\documentclass[prl]{revtex4}
\usepackage{graphicx}
\usepackage{bm}
\usepackage{amssymb}
\usepackage{amsmath}

\begin{document}

\title{Superconductivity of electron-hole pairs in a
bilayer graphene system in a quantizing
magnetic field}
\author{ D. V. Fil and L. Yu. Kravchenko\\
\emph{Institute of Single Crystals of the National Academy of
Sciences of Ukraine, Lenin av. 60, Kharkov 61001, Ukraine}}
%\date{ }

\maketitle

\begin{center}
Abstract
\end{center}
The state with a spontaneous interlayer phase coherence in a
graphene based  bilayer quantum Hall system is studied. This state
can be considered as a gas of superfluid electron-hole pairs with
the components of the pair belonging to different layers.
Superfluid flux of such pairs is equivalent to two electrical
supercurrents in the layers. It is shown that the state with the
interlayer phase coherence emerges in the graphene system if a
certain imbalance of the Landau level filling factors  of the
layers is created. We obtain the temperature of transition into
the superfluid state, the maximum interlayer distance at which the
phase coherence is possible, and the critical values of the
supercurrent. The advantages of use of graphene systems instead of
GaAs heterostructures for the realization of the bilayer
electron-hole superconductivity is discussed.


\section{Introduction}

The experimental discovery of graphene \cite{1} served as a
stimulus for performing large-scale investigations of the
electronic properties of conductors with a Dirac dispersion law
for charge carriers. On the one hand such investigations are
certainly fundamental. On the other hand the uniqueness of the
properties of graphene makes it an extremely interesting material
from the standpoint of different practical applications.

In the present article we shall examine the prospects for using
graphene to obtain superconductivity of bound electron-hole pairs
in a system of two two-dimensional conductors separated by a
dielectric. The possibility of such superconductivity was
suggested quite a long time ago \cite{2,3}. The system considered
was a sandwich consisting of a two-dimensional \textit{n}-type
semiconductor -- dielectric -- two-dimensional \textit{p}-type
semiconductor. Somewhat later it was determined \cite{4,5} that a
similar phenomenon can be obtained in a semiconductor
heterostructure with two quantum wells (quasi-two-dimensional
electronic layers). If such a system is placed in the quantizing
magnetic field oriented perpendicular to the layers (a quantum
Hall effect regime is realized) and the total filling factor of
the Landau levels is 1 ($\nu_1+\nu_2=1$, where $\nu_i$ is the
filling factor in a layer), pairing of electrons in one layer with
holes (unfilled states in the lowest Landau level) in the other
layer occurs. These pairs are bosons; as temperature decreases,
such a pair gas can transition into a superfluid state. Because
pair gas is two-dimensional the transition will occur by the
Berezinskii-Kosterlitz-Thouless mechanism. Since the electron and
hole components of the pairs belong to different layers, the
superfluid flow of the electrically neutral pairs is equivalent to
two electric supercurrents flowing in opposite directions in
neighboring layers.

The analogy with a quantum ferromagnet is widely used to describe
such a state (which is called a state with spontaneous interlayer
phase coherence)\cite{6,7,8,9,10}. This analogy makes it possible
to study, within the framework of the same approach, the case of
zero imbalance of the filling factors of the layers
($\nu_1=\nu_2=1/2$) as well as a situation with nonzero imbalance
($\nu_1=\nu$, $\nu_2=1-\nu$, $\nu \ne 1/2$), which arises, for
example, because of the external electrostatic field of a gate.
For a very large imbalance ($\nu\ll 1$) the system can be
described as a rarefied gas of bosons (magnetoexcitons) with a
dipole-dipole interaction. Such an approach was developed in Refs.
\cite{11,12,13,14,15}. We note that experimental investigations of
two-layer electron-hole superconductivity have been performed
primarily on quantum Hall system \cite{16,17,18,19,20,22,23,24}.

The superconductivity of electron-hole pairs in a
graphene-dielectric-graphene system was studied in Refs.
\cite{27,28,29,30,31,32,33}. In Ref. \cite{27,28,29,30} and
\cite{32} the possibility of such pairing was studied in the
absence of a magnetic field perpendicular to the layers (i.e. not
in the quantum Hall effect regime). In the absence of a magnetic
field the condition for BCS pairing of electrons and holes is that
their Fermi surfaces must coincide. For semiconductors with a
quadratic dispersion law for the charge carriers the last
condition requires that the electron and hole effective masses be
the same. For a Dirac carrier spectrum (which obtains in graphene)
the required condition on the Fermi surface is satisfied
automatically. Neglecting screening the theory predicts a very
high (hundreds of degrees) temperature for the transition into the
superconducting state in such a system \cite{27,32}. Taking
account of screening the estimate for the transition temperature
is much less optimistic and lies in the millikelvin range
\cite{30,31}.

In a quantum Hall system the screening effects should not be so
strong, which gives hope of attaining high temperatures for the
transition into the superconducting state. In addition, the
quantum Hall effect itself in graphene is observed at high
temperatures (right up to room temperatures) \cite{rtg}. The
question of the superconductivity of a gas of magnetoexcitons in a
two-dimensional graphene system was examined in Refs. \cite{31}
and \cite{33}. The approach of Refs. \cite{31,33} can be use only
for low magnetoexciton densities ($\nu\ll 1$). But the low-density
case is not optimal from the standpoint of reaching high
transition temperatures. More likely, the maximum critical
temperature will obtain for half-filling of the Landau levels in
each layer. The present article is devoted to analyzing this
problem.

\section{State with interlayer phase coherence in a
bilayer graphene system}

The starting point for studying electron-hole pairing in a quantum
Hall system is the Coulomb interaction Hamiltonian written in the
lowest (active) Landau level approximation. Consequently, we shall
begin by obtaining the desired Hamiltonian for the graphene
system.

To describe the properties of the electronic subsystem of graphene
in a magnetic field we shall follow the approach presented in
detail in the review  \cite{34} (see also Ref. \cite{35}).
Graphene possesses a honeycombed crystal structure, which can be
represented as two simple triangular lattices A and B inserted
into one another. The distance a between the nearest carbon atoms
(which belong to different sublattices) is 1.42 \AA. In the
tight-binding approximation and taking account of tunneling only
between the nearest sites the Hamiltonian of the system has the
form
\begin{equation}\label{1}
H=-t\sum_{\langle i,j \rangle, \sigma}
(a_{i,\sigma}^+b_{j,\sigma}+h.c),
\end{equation}
where $a^+_i$, $b^+_i$ ($a_i$, $b_i$) are, respectively, operators
creating (annihilating) electrons on the i-th sites of the
sublattices A and B, $t\approx 2.8$ eV is the tunneling amplitude,
and $\sigma$ is the spin index.

Having written the Hamiltonian (\ref{1}) in the momentum
approximation it is easy to verify that the energy band is divided
into two sub-bands which possess only two nonequivalent points of
contiguity ${\bf K}=(2\pi/3a,2\pi/3\sqrt{3}a)$ and ${\bf
K}'=(2\pi/3a,-2\pi/3\sqrt{3}a)$ (the $x$ axis is directed along
the line connecting any pair of nearest-neighbor sites). In
undoped graphene (containing one free electron per site) the Fermi
level passes through the point of contiguity of these sub-bands.
Correspondingly, the sub-bands can be interpreted as electron and
hole bands. In connection with such a structure of the spectrum
the low-energy excitations in undoped and weakly doped graphene
can be described by adding a pseudospin index $\alpha=\pm 1$,
corresponding to states whose quasimomenta lie close to ${\bf K}$
and ${\bf K}'$. For low energies the Hamiltonian is diagonal with
respect to the pseudospin indices. It is also assumed that there
are no interactions which destroy the diagonality with respect to
spin. The characteristic quantum numbers of such a Hamiltonian
possess a definite spin and pseudospin.

For such states the Schrodinger equation written in the coordinate
representation has the form
\begin{equation}\label{2}
 - i \hbar v_F \left(%
\begin{array}{cc}
  0 & \frac{\partial}{\partial x}-i \alpha \frac{\partial}{\partial y}\\
  \frac{\partial}{\partial x}+i \alpha \frac{\partial}{\partial y} & 0 \\
\end{array}%
\right)
\left(%
\begin{array}{c}
  \Psi_A \\
  \Psi_B \\
\end{array}%
\right)_{\alpha,\sigma}= E \left(%
\begin{array}{c}
  \Psi_A \\
  \Psi_B \\
\end{array}%
\right)_{\alpha,\sigma}
\end{equation}
where $v_F=3ta/2 \hbar$. For graphene the indicated parameter
equals $\approx 10^8$ cm/ s. An electronic state is a spinor whose
components correspond to the sublattices A and B. Switching in Eq.
(\ref{2}) to the momentum representation we find the spectrum in
the form $E=\pm \hbar v_F k$ ($k$ is the modulus of the wave
vector, measured from the point ${\bf K}$ or ${\bf K}'$).
Evidently, $v_F$ is the velocity of the electrons on the Fermi
surface. For fermions with such a spectrum this velocity does not
depend on the carrier concentration.

In a magnetic field perpendicular to graphene Schrodinger's
equation (\ref{2}) assumes the form
\begin{equation}\label{3}
 - i \hbar v_F \left(%
\begin{array}{cc}
  0 & \frac{\partial}{\partial x}+\frac{\alpha e B x}{\hbar c} -i \alpha \frac{\partial}{\partial y}\\
  \frac{\partial}{\partial x}-\frac{\alpha e B x}{\hbar c} +i \alpha \frac{\partial}{\partial y} & 0 \\
\end{array}%
\right)
\left(%
\begin{array}{c}
  \Psi_A \\
  \Psi_B \\
\end{array}%
\right)= E \left(%
\begin{array}{c}
  \Psi_A \\
  \Psi_B \\
\end{array}%
\right).
\end{equation}
The solution of Eq. (\ref{3}) gives the Landau energy levels in
graphene: $E_0=0$, $E_{\pm N}=\pm (\hbar v_F/\ell)\sqrt{2N}$,
where $\ell=\sqrt{\hbar c/eB}$ is the magnetic length and
$N=1,2,\ldots$ In Eq. (\ref{3}) we neglected the Zeeman terms,
since for reasonable values of the magnetic field the Zeeman
splitting is much less than the spacing between the Landau levels.
The eigenfunctions ${\bf \Psi}_{N,\alpha,k}$ corresponding to the
zeroth and $\pm N$-th Landau levels have the following form:
\begin{equation}\label{4}
{\bf \Psi}_{0,-1,k}(x,y)=\frac{e^{-i k
y}}{\pi^{1/4}\sqrt{\ell L_y}}e^{-\frac{(x-X)^2}{2\ell^2}}\left(%
\begin{array}{c}
  1 \\
  0 \\
\end{array}%
\right), \quad {\bf \Psi}_{0,+1,k}(x,y)=\frac{e^{-i k
y}}{\pi^{1/4}\sqrt{\ell L_y}}e^{-\frac{(x-X)^2}{2\ell^2}}\left(%
\begin{array}{c}
  0 \\
  1 \\
\end{array}%
\right),
\end{equation}
\begin{eqnarray}\label{5}
  {\bf \Psi}_{\pm N,-1,k}(x,y)=\frac{e^{-i k
y}}{\pi^{1/4}\sqrt{\ell L_y}\sqrt{2^{N+1}N!}}e^{-\frac{(x-X)^2}{2\ell^2}}\left(%
\begin{array}{c}
  \mp H_N\left(\frac{x-X}{\ell}\right) \\
  i \sqrt{2N} H_{N-1}\left(\frac{x-X}{\ell}\right)\\
\end{array}%
\right), \cr {\bf \Psi}_{\pm N,+1}(x,y)=\frac{e^{-i k
y}}{\pi^{1/4}\sqrt{\ell L_y}\sqrt{2^{N+1}N!}}e^{-\frac{(x-X)^2}{2\ell^2}}\left(%
\begin{array}{c}
  i \sqrt{2N} H_{N-1}\left(\frac{x-X}{\ell}\right) \\
  \mp H_N\left(\frac{x-X}{\ell}\right)\\
\end{array}%
\right),
\end{eqnarray}
where $X=k\ell^2$, and $H_N(x)$ is a Hermite polynomial. The
degeneracy of a Landau level is $4S/(2\pi \ell^2)$, where $S=L_x
L_y$ is the area of the layer. The factor of 4 is due to the
two-fold degeneracy with respect to both spin index and the index
$\alpha$. We introduce the filling factor $\nu=2\pi\ell^2 n$,
where $n$ is the concentration of filled states. A completely
filled level corresponds to $\nu=4$. In undoped graphene the
chemical potential corresponds to the zeroth Landau level, so that
the negative values are completely filled, the positive values are
empty, and the zeroth level is half-filled (its filling factor is
2).

Let us consider a system of two graphene layers separated by a
dielectric layer with thickness $d$ and permittivity
$\varepsilon$. To avoid complications which are not of a
fundamental nature we shall assume that the system is located
inside a dielectric matrix with the same permittivity
$\varepsilon$. The Coulomb interaction Hamiltonian for such a
system has the form
\begin{equation}\label{6}
    H_C=\frac{1}{2}\sum_{i,i'}\int d^2 r d^2 r'
    V_{i,i'}(|{\bf r}-{\bf r}'|) \hat {\rho}_{i} ({\bf r})
    \hat{\rho}_{i'} ({\bf r}').
\end{equation}
where $V_{i,i'}(r)=e^2/(\varepsilon \sqrt{r^2+d^2(i-i')^2})$ is
the Coulomb potential, $$\hat {\rho}_{i} ({\bf
r})=\sum_{\alpha,\sigma}\hat{\Psi}^+_{i,\alpha,\sigma}({\bf
r})\hat{\Psi}_{i,\alpha,\sigma}({\bf r})$$  is the electron
density operator, $i=1,2$ is the index of the layer, and
$\hat{\Psi}^+_{i,\alpha,\sigma}({\bf r})$ and
$\hat{\Psi}_{i,\alpha,\sigma}({\bf r})$ are operators creating and
annihilating at the point ${\bf r}$ an electron with prescribed
spin and pseudospin.

We shall represent the operators $\hat{\Psi}^+$, $\hat{\Psi}$ in
terms of the operators (\ref{4}) and (\ref{5}) creating and
annihilating electrons in the quantum states, respectively. If the
separation of the Landau levels is much greater than the Coulomb
energy $e^2/(\varepsilon \ell)$, then it is sufficient to retain
in this expansion only a single active (partially filled) Landau
level
\begin{equation}\label{7}
\hat{\Psi}_{i,\alpha,\sigma}(x,y)= \sum_{k}
\Psi_{\lambda,\alpha,k}(x,y) a_{i,k,\alpha,\sigma}, \quad
\hat{\Psi}^+_{i,\alpha,\sigma}(x,y)= \sum_{k}
\Psi_{\lambda,\alpha,k}^+(x,y) a_{i,k,\alpha,\sigma}^+,
\end{equation}
where $\lambda$ is the number of the active level,
$\Psi_{\lambda,\alpha,k}^+$ are two-component vector-matrices
determined by Eqs. (\ref{4}) and (\ref{5}),
$\Psi_{\lambda,\alpha,k}^+$ are the hermitian-conjugate matrices,
and $a_{i,k,\alpha,\sigma}^+$ and $a_{i,k,\alpha,\sigma}$ are
operators creating and annihilating electrons in a state with the
corresponding quantum numbers in the level $\lambda$.

In this approximation the Hamiltonian (\ref{6}) written in the
Fourier representation has the form
\begin{equation}\label{8}
    H_C=\frac{1}{2S}\sum_{i,i'}\sum_{\bf q}
    V_{i,i'}(q)\hat{\rho}_i({\bf q})\hat{\rho}_{i'}(-{\bf q}),
\end{equation}
where $V_{i,i'}(q)=(2\pi e^2/\varepsilon q) \exp(-q d|i-i'|)$ is
the Fourier component of the Coulomb potential. For $\lambda=0$
the Fourier component of the electron density operator is
\begin{equation}\label{9}
\hat{\rho}_i({\bf q})=\sum_{\alpha,\sigma} \sum_k
a^+_{i,k+q_y/2,\alpha,\sigma} a_{i,k-q_y/2,\alpha,\sigma}
\exp\left(-iq_x k \ell^2-\frac{q^2\ell^2}{4}\right).
\end{equation}
For $\lambda\ne 0$
\begin{equation}\label{10}
\hat{\rho}_i({\bf q})=\sum_{\alpha,\sigma} \sum_k
a^+_{i,k+q_y/2,\alpha,\sigma} a_{i,k-q_y/2,\alpha,\sigma}
\exp\left(-iq_x k
\ell^2-\frac{q^2\ell^2}{4}\right)\frac{L_{|\lambda|}\left(\frac{q^2\ell^2}{2}\right)+
L_{|\lambda|-1}\left(\frac{q^2\ell^2}{2}\right)}{2},
\end{equation}
where $L_\lambda(x)$ is a Laguerre polynomial.

If the system is composed of two undoped graphene layers, then in
each layer the half-filled zeroth Landau level will be active. If
an electrostatic field is applied in a direction perpendicular to
the layers, then an imbalance of the filling factor can arise,
i.e. the filling factors of the layers will become equal to
$\nu_1=2+\tilde{\nu}$ and $\nu_2=2-\tilde{\nu}$
($0<\tilde{\nu}\leq 2$). In stronger fields the active levels will
become $\lambda=+1$  in layer 1 and $\lambda=-1$ in the layer 2,
then $+2$ and $-2$, and so on.

We shall examine first the case where the zeroth levels are the
active levels. The Coulomb interaction Hamiltonian differs from
the Hamiltonian of the usual bilayer quantum Hall system with
total filling factor $\nu=1$ only by the fact that it contains an
interaction of not one but four components, corresponding to the
quantum numbers $(\alpha=\pm 1, \sigma=\uparrow, \downarrow)$. We
shall enumerate these sets of quantum numbers by the index
$\beta=1,2,3,4$.

By analogy with the single-component case \cite{6,7} we write the
trial multi-particle wave function of the system in the form
\begin{equation}\label{11}
 |\Psi\rangle =
 \prod_{k}\prod_{\beta}\left(\cos\frac{\theta_{\beta}}{2}
 a^+_{1,k\beta}+ e^{i \varphi_\beta}\sin \frac{\theta_\beta}{2}
 a^+_{2,k\beta}\right)|0\rangle.
\end{equation}
The function (\ref{11}) describes a state where interlayer phase
coherence arises for each component. It is easily shown (see, for
example, Refs. \cite{36,37}) that the state (\ref{11}) can be
represented in the form of a BCS wave function describing the
pairing of electrons of one layer with holes of the other layer
with the paired electrons and holes corresponding to the same
state $\beta$. The parameter $\theta_\beta$ is related with the
magnitude of the imbalance for this component by the relation
$\tilde{\nu}_\beta=\cos \theta_\beta/2$; the filling factors for
this component are $\nu_{1(2)\beta}=1/2\pm \tilde{\nu}_\beta$, and
the total imbalance is
$$\tilde{\nu}=\sum_\beta \tilde{\nu}_\beta$$
The energy of the system in the state (\ref{11}) is given by the
expression $E=\langle \Psi|H_C|\Psi\rangle+ E_\textrm{g} + E_{\rm
bg}$, where $E_\textrm{g}$ is the interaction energy of the
electrons with the electrostatic field produced by the external
gate, and $E_\textrm{bg}$ takes account of the Coulomb interaction
with the positive core. A direct calculation (see, for example,
Ref. \cite{38}) gives
\begin{equation}\label{12}
    E=\frac{S}{2\pi \ell^2}\left(W
    \left(\sum_\beta\tilde{\nu}_\beta\right)^2
    -J_0\left(1+\sum_\beta\tilde{\nu}_\beta^2\right)-J_1
    \left(1-\sum_\beta\tilde{\nu}_\beta^2\right)-  eV
    \sum_\beta\tilde{\nu}_\beta\right),
\end{equation}
\begin{equation}\label{12a}
W=\frac{e^2 d}{\varepsilon \ell^2}, \quad
J_0=\sqrt{\frac{\pi}{2}}\frac{e^2}{\varepsilon\ell}, \quad J_1=J_0
\exp\left(\frac{d^2}{2\ell^2}\right){\rm
erfc}\left(\frac{d}{\sqrt{2}\ell}\right),
\end{equation}
where ${\rm erfc}(x)$ is the complementary error function and V is
the potential difference produced between the layers by the gate
field.

The first term in Eq. (\ref{12}) is identical to the energy of a
plane capacitor and represents the contribution of the direct
Coulomb interaction to the energy. The second and third terms
contribute to the exchange interaction inside and be tween the
layers, respectively. We note that $J_0>J_1$, i.e. the in-layer
exchange energy constant is greater than the interlayer exchange
energy constant.

To find the parameters $\theta_\beta$ appearing in the trial
function (\ref{11}) it is necessary to find $\tilde{\nu}_\beta$
corresponding to the minimum of the energy (\ref{12}) taking
account of the constraint $|\tilde{\nu}_\beta|\leq 1/2$ (which
follows from the definition of $\tilde{\nu}_\beta$).

In a nongraphene bilayer quantum Hall system with $\nu_1+\nu_2=1$
there is only one component, and the energy minimum in the absence
of a gate field corresponds to $\tilde{\nu}=0$, i.e.
$\nu_1=\nu_2=1/2$. The situation is more complicated in a
multi-component system. For $J_0>J_1$ the part quadratic in
$\tilde{\nu}_\beta$ in Eq. (\ref{12}) is not positive-definite.
Moreover, it is not positive-definite with one and two
$\tilde{\nu}_\beta$ fixed. Therefore the minimum of the energy
(\ref{12}) is attained at the boundary of the range of
$\tilde{\nu}_\beta$, and in addition at least three of the four
$\tilde{\nu}_\beta$ must assume limiting values, equal $\pm 1/2$.
For $V=0$ the minimum is reached for
$$\sum_\beta\tilde{\nu}_\beta=0,$$ but when two $\tilde{\nu}_{\beta_i}=+1/2$ and the
remaining other two $\tilde{\nu}_{\beta_{i'}}=-1/2$. Physically,
this result is understandable. Since $J_0>J_1$, a gain in the
energy exchange is attained for maximum imbalance of a given
component. In a one-component system a direct exchange interaction
impedes this. In a multi-component system the latter effect is
absent, since the magnitudes of the imbalance of the components
can be of opposite sign. The values $\tilde{\nu}_\beta=\pm 1/2$
correspond to $\theta_\beta=0,\pi$ which means absence of
interlayer phase coherence (i.e. absence of a superconducting
state). We note that this situation is analogous to the one
arising in a nongraphene bilayer quantum Hall system with total
filling factor $\nu_1+\nu_2=2$ \cite{39}.

We shall show that for $V\ne 0$ it becomes possible for a state
with interlayer phase coherence to appear. For definiteness, we
shall assume that $V>0$. We shall also assume that the parameters
determined by Eqs. (\ref{12a}) satisfy the inequality
$W-J_0+J_1>0$. Analysis of Eq. (\ref{12}) shows that for $eV\leq
J_0-J_1$ an energy minimum corresponds to the same state as for
$V=0$. If the potential difference satisfies the condition
\begin{equation}\label{13}
    J_0-J_1<eV<2W-J_0+J_1,
\end{equation}
then the minimum energy is attained in the state
\begin{equation}\label{14}
    \tilde{\nu}_1=\tilde{\nu}_2=\frac{1}{2},\quad
    \tilde{\nu}_3=\frac{eV-W}{2(W-J_0+J_1)}, \quad
    \tilde{\nu}_4=-\frac{1}{2}.
\end{equation}
The state (\ref{14}) is degenerate with respect to transposition
of the indices $\beta$. Most likely, such degeneracy is removed as
a result of weaker interactions which are not taken into account
in Eq. (\ref{12}). This is not essential for our analysis. In the
parameter range
\begin{equation}\label{15}
    2W-J_0+J_1\leq eV \leq 2W+J_0-J_1
\end{equation}
the minimum corresponds to the state
\begin{equation}\label{16}
    \tilde{\nu}_1=\tilde{\nu}_2=\tilde{\nu}_3=\frac{1}{2},\quad
    \tilde{\nu}_4=-\frac{1}{2}.
\end{equation}
If $V$ lies in the range
\begin{equation}\label{17}
    2W+J_0-J_1\leq e V \leq 4W-J_0+J_1,
\end{equation}
then the minimum is reached when
\begin{equation}\label{18}
    \tilde{\nu}_1=\tilde{\nu}_2=\tilde{\nu}_3=\frac{1}{2},\quad
    \tilde{\nu}_4=\frac{eV-3W}{2(W-J_0+J_1)}.
\end{equation}
As $V$ increases further, the zeroth Landau level will become
inactive: it will be completely filled in layer 1 and empty in
layer 2.

In the next section we shall see that the superfluid density and,
correspondingly, the transition temperature are proportional to
$\sqrt{\nu_\beta(1-\nu_\beta)}=\sqrt{1/4-\tilde{\nu}_\beta^2}$,
i.e. different from zero for $\tilde{\nu}_\beta\ne \pm 1/2$. As
follows from the expressions presented above, pairing will arise
when the external potential difference lies in the range
(\ref{13}) and (\ref{17}). Then the maximum superfluid density
(corresponding to $\tilde{\nu}_\beta=0$) will correspond to $eV=W$
and $eV=3W$. To estimate the required field strength in the gate
we note that the condition $eV=W$ corresponds to field strength
$E\approx  2\cdot 10^4 \varepsilon^{-1} B$ V/cm, where $B$ is
taken in T.

We also note that for $eV=W$ the filling factors of the layers are
$\nu_1=5/2$ and $\nu_2=3/2$, and for $eV=3W$ they assume the
values $\nu_1=7/2$ and $\nu_2=1/2$. In both cases only one
component has a nonzero imbalance between the layers, and it is
this (active) component that is responsible for the pairing of
electrons and holes.

As $V$ increases further, the electrons will transition from the
level $\lambda=-1$ of layer 2 to the level $\lambda=+1$ of layer
1, and pairing of electrons in the $\lambda=+1$ level with holes
in the $\lambda=-1$ level will become possible. According to Eq.
(\ref{10}), the expressions for the Fourier components of the
density operator for the levels $\lambda=\pm 1$ have the form
\begin{equation}\label{19} \hat{\rho}_i({\bf
q})=\sum_{\alpha,\sigma} \sum_k a^+_{i,k+q_y/2,\alpha,\sigma}
a_{i,k-q_y/2,\alpha,\sigma} \left(1-\frac{q^2\ell^2}{4}\right)
\exp\left(-i q_x k \ell^2-\frac{q^2\ell^2}{4}\right).
\end{equation}
The state with pairing of electrons in level $\lambda=+1$ with
holes in level $\lambda=-1$ can be described by a wave function of
the form ({\ref{11}) in which the operators creating electrons in
the layer 1 belong to the Landau level $\lambda=+1$ and the
operators creating electrons in the layer 2 belong to the level
$\lambda=-1$.

The Coulomb interaction energy in such a state, taking account of
the imbalance of the components of the level $\lambda=0$, assumes
the form
\begin{eqnarray}\label{20}
    E=\frac{S}{2\pi \ell^2}\Bigg(W
    \left(2+\sum_\gamma\left(\frac{1}{2}+\tilde{\nu}_\gamma\right)\right)^2 -2 J_0
    - J_2\left(1+\sum_\gamma\tilde{\nu}_\gamma^2\right)\cr -J_3
    \left(1-\sum_\gamma\tilde{\nu}_\gamma^2\right)- 2eV -
    (eV-\Omega)
    \sum_\gamma\left(\frac{1}{2}+\tilde{\nu}_\gamma\right)\Bigg),
\end{eqnarray}
where
\begin{equation}\label{21}
    J_2=\frac{11}{16}\sqrt{\frac{\pi}{2}}\frac{e^2}{\varepsilon\ell},\quad
    J_3=J_2
    \left[\frac{11-2\tilde{d}^2+\tilde{d}^4}{11}
    \exp\left(\frac{\tilde{d}^2}{2}\right){\rm
    erfc}\left(\frac{\tilde{d}}{\sqrt{2}}\right)
    +\frac{2\tilde{d}(3-\tilde{d}^2)}{11\sqrt{2\pi}}\right]
\end{equation}
are the in-layer and interlayer exchange interaction constants for
the levels $|\lambda|=1$ ($\tilde{d}=d/\ell$). As in the case of
an active zeroth level, the in-layer exchange constant $J_2$ is
greater than the interlayer exchange constant $J_3$. In Eq.
(\ref{20}) $\Omega=2\sqrt{2}\hbar v_F/\ell$ is the splitting
between the levels $\lambda=+1$ and $\lambda=-1$, $\gamma$ denote
the same quantum numbers as $\beta$ but for the levels
$|\lambda|=1$, and $\tilde{\nu}_\gamma$ determine the imbalance of
the components of $\gamma$.

We find from Eq. (\ref{20}) that for
\begin{equation}\label{22}
    4W+\Omega+J_2-J_3\leq eV \leq 6W+\Omega-J_2+J_3
\end{equation}
the minimum energy corresponds to the state
\begin{equation}\label{23-1}
\tilde{\nu}_1=\frac{eV-\Omega-5W}{2(W-J_2+J_3)},\quad
\tilde{\nu}_2=\tilde{\nu}_3=\tilde{\nu}_4=-\frac{1}{2},
\end{equation}
i.e. the inequalities (\ref{22}) determine the next range of $V$
where a state with interlayer phase coherence can arise but this
time between the electrons belonging to the levels $\lambda=+1$
and $\lambda=-1$. Specifically, for $eV=\Omega+5W$ the component
in which such coherence will arise will have zero imbalance, i.e.
the maximum superfluid density will be reached.

\section{Critical parameters for electron-hole
superconductivity in graphene}

As shown in the preceding section, a state with interlayer phase
coherence can arise in a bilayer graphene system in a quantizing
magnetic field and an electric field perpendicular to the layers.
We established that even though four components are present in a
Landau level in graphene, only one component can be active.
Depending on the magnitude of the gate field interlayer phase
coherence can arise between the electrons in the zeroth Landau
level or between the electrons in the $\lambda=\pm 1$ levels. In
the first case the critical parameters of electron-hole
superconductivity in graphene will be described by the same
equations as in bilayer quantum Hall systems in GaAs
heterostructures. In the second case the equations will be
somewhat different. In the present section we shall compare two
such cases. We shall use for the analysis the approach of Ref.
\cite{10}, which was further elaborated in Ref. \cite{40}, as well
as in the present work.

We shall examine a state where $V$ gives interlayer phase
coherence of one of the components, and in which the parameters
$\theta$ and $\varphi$ for this component depend on $k$:
\begin{equation}\label{23}
 |\Psi\rangle =
 \prod_{k}\left(\cos\frac{\theta_k}{2}
 a^+_{1,k}+ e^{i \varphi_k}\sin \frac{\theta_k}{2}
 a^+_{2,k}\right)|0\rangle.
\end{equation}
Here and below the creation and annihilation operators as well as
the functions $\theta_k$ and $\varphi_k$ refer to the active
component. The quantum index $k$ determines the coordinate $x$ of
the center of the electron orbit: $X=k \ell^2$. Consequently, the
function (\ref{23}) describes a state in which the order parameter
$\langle\Psi |a_{1X}^+a_{2X}|\Psi\rangle=(1/2)\sin \theta_X
e^{i\varphi_X}$ for electron-hole pairing varies along the $x$
axis.

The energy of the system in the state (\ref{23}) is
\begin{eqnarray} \label{24}
 E =E_0 -{e\tilde{V}} \sum_{X}  \cos \theta_X+\frac {1} {2 L_{y}} \sum _ {X, X '} \Big \{\left [H
(X-X ') - F_\textrm{S} (X-X ') \right] \cos \theta_X \cos \theta _
{X '} \cr
 - F_\textrm{D} (X-X ') \sin\theta_X \sin\theta _ {X '}
\cos (\varphi_X - \varphi _ {X '}) \Big \},
\end{eqnarray}
where $\tilde{V}=V-V_0$ is the potential difference corresponding
to zero imbalance of this component and $E_0$ is the Coulomb
energy of the inactive components. To abbreviate the notations
appearing in Eq. (\ref{23}) we shall express the functions of
$X-X'$ in terms of the relation $A(X)=\ell^2
 \int d q {\mathcal A}(q) e^{iqX}$, where
$A(X)=H(X),F_\textrm{S}(X)$ and $F_\textrm{D}(X)$ and the explicit
form the Fourier component of these quantities is as follows:
\begin{equation} \label{62}
    \mathcal {H} (q) = \frac {e^2} {2 \varepsilon l^2} \; e ^ {-\frac {q^2 l^2} {2}}
    \frac {1 - e ^ {-d |q |}} {|\, q |} f_\lambda(q\ell),
\end{equation}
\begin{equation} \label{63}
    \mathcal {F} _\mathrm{S} (q) = \frac {e^2} {2 \varepsilon \ell} \int_0 ^ {\infty}
    d k \, e ^ {-\frac {k^2
    } {2}} J_0 (k q \ell) f_\lambda(k), \quad
\end{equation}
\begin{equation} \label{64}
    \mathcal {F} _\mathrm{D} (q) = \frac {e^2} {2\varepsilon \ell} \int_0 ^ {\infty}
    d k \; e ^ {-\frac {k^2
    } {2}} J_0 (k q \ell) \; e ^ {-k \tilde{d}} f_\lambda(k).
\end{equation}
In Eqs. (\ref{62})-(\ref{64}) $J_0(q)$ is the Bessel function of
order zero, and
\begin{equation}\label{64a}
    f_\lambda(k)=\begin{cases} 1&\textrm{for }  \lambda =0
    \\\left(1-\frac{k^2}{4}\right)^2& \textrm{for }  |\lambda| = 1
    \end{cases}
\end{equation}
(the active levels are $\lambda=0$ or $\lambda=\pm 1$).

In the uniform sate with nonzero flux of electron-hole pairs in
the $x$ direction the phase of the order parameter is linear in
$X$ ($\varphi_X=Q X$) and $\theta_X=\theta_0$ is independent of
$X$. The energy of this state is
\begin{equation} \label{65}
    E_\textrm{mf} = E_0-\frac{e \tilde{V} S \cos \theta_0}{2\pi\ell^2}+
    \frac {S} {4 \pi l^2} \Bigg(\left [\mathcal {H} (0) - \mathcal {F} _\textrm{S}
(0) \right] \cos^2 \theta_0 - \mathcal {F} _\textrm{D} (Q)
\sin^2\theta_0 \Bigg).
\end{equation}
Since the imbalance $\tilde{\nu}_\beta=\cos{\theta_0}/2$ and
$2\mathcal {H} (0)=W$, $2\mathcal {F} _\textrm{S}(0)=J_0\ (J_2)$,
and $2\mathcal {F} _\textrm{D}(0)=J_1 \ (J_3)$ (for $\lambda=0\
(\pm 1)$ respectively), it is evident that there is complete
correspondence between the expression (\ref{65}) at $Q=0$ and the
expressions (\ref{12}) and (\ref{20}).

For small $Q$ the energy (\ref{65}) can be represented as
\begin{equation}\label{66}
    E=S(\textrm{const}+\frac{1}{2}\rho_{s0} Q^2),
\end{equation}
where
\begin{equation}\label{67-0}
\rho_{s0}= \sin^2 \theta_0 \frac{E_c}{16 \pi}
\tilde{\rho}\left(\frac{d}{\ell}\right),
\end{equation}
$E_c=e^2/\varepsilon \ell$ is the Coulomb energy, and the function
\begin{equation}\label{67}
   \tilde{\rho}(x)=\begin{cases} \sqrt{\frac{\pi}{2}}e^{\frac{x^2}{2}}{\rm erfc}
   \left(\frac{x}{\sqrt{2}}\right)(1+x^2)-x,
   &\textrm{for }  \lambda =0 \\
   \sqrt{\frac{\pi}{2}}e^{\frac{x^2}{2}}{\rm erfc}
   \left(\frac{x}{\sqrt{2}}\right)\frac{7+13 x^2+7x^4+x^6}{16}
   -\frac{x(3+x^2)^2}{16}&\textrm{for }  |\lambda| = 1\end{cases}
\end{equation}
gives the dependence of this quantity on the distance between the
layers.

The quantity $\rho_{s0}$ is called the superfluid stiffness. More
precisely, the expression  (\ref{67-0}) gives the value of this
quantity at $T=0$ in the mean-field approximation. A plot of the
function $\rho_{s0}(d/\ell)$ is displayed in Fig. \ref{f1}. The
quantity $\pi \rho_{s0}/2$ gives an estimate of the temperature of
the transition into the superfluid state. This estimate neglects
the temperature correction to $\rho_s$ and is valid in a bounded
interval of $d$. We shall discuss this question in greater detail
below.

\begin{figure}
\begin{center}
\includegraphics[width=6cm]{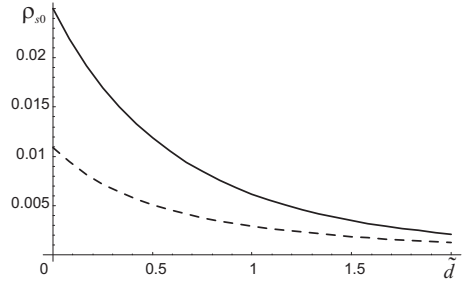}
\end{center}
\caption{Dependence of the superfluid stiffness (in the units $E_c
\sin^2 \theta_0$) on the distance between the layers
($\tilde{d}=d/\ell$). Solid line -- $\lambda=0$ (pairing on the
zeroth Landau level), dashed line -- $|\lambda|=1$ (pairing on the
$\pm 1$ levels).} \label{f1}
\end{figure}

To find the critical current and the transition temperature it is
necessary to obtain the spectrum of the collective modes. We shall
use the approach of Refs. \cite{10} and \cite{40}, which is based
on the quantization of the energy of small fluctuations of
$\theta_X$ and $\varphi_X$. First we shall require the spectrum
for wave vectors $q$ directed along the gradient of the phase
$\varphi$ (which was obtained in Ref. \cite{40}). The general case
is analyzed in the Appendix. According to Ref. \cite{40}, the
expression for the spectrum can be written in a form that is
formally identical to the expression for the Bogolyubov spectrum
of quasiparticles in a moving condensate:
\begin{equation} \label{39-1}
E (q) = \sqrt {\epsilon  (\epsilon   +2 \gamma )} + \hbar \, q v.
\end{equation}

The following notations have been introduced in Eq. (\ref{39-1})
\begin{equation} \label{40-1} \epsilon =2\mathcal {F} _{D}
(Q) - \mathcal {F} _{D} (q+Q) - \mathcal {F} _{D} (q-Q),
\end{equation}
\begin{equation} \label{40-2}
\gamma  =  \sin^2\theta_0 \left [\mathcal {H} (q) - \mathcal {F}
_{S} (q) + \frac{\mathcal {F} _{D} (q+Q)+ \mathcal {F} _{D}
(q-Q)}{2}\right],
\end{equation}
\begin{equation} \label{40-3}
v = \frac {\mathcal {F} _\mathrm{D} (q+Q) - \mathcal {F}
_\mathrm{D} (q - Q)} {\hbar q} \cos \theta_0.
\end{equation}
The quantity (\ref{40-1}) is the kinetic energy of pairs. In the
limits $q\to 0$ and $Q \to 0$ the expression (\ref{40-1}) reduces
to $\epsilon=\hbar^2 q^2/2M$, where $M$ is the magnetic mass of a
pair
\begin{equation}\label{40-4}
    M=\frac{2 \hbar^2 \varepsilon }{e^2 \ell }
    \frac{1}{\tilde{\rho}(d/l)},
\end{equation}
and $\tilde{\rho}(x)$ is determined by Eq. (\ref{67}). Since the
mass $M$ is inversely proportional to $\rho_s$, according to Fig.
1 it increases with $d$, and the magnetic mass for the active
levels $\lambda=\pm 1$ is greater than that for the active level
$\lambda=0$.

The expression (\ref{40-2}) in the limits $q\to 0$ and $Q \to 0$
goes to a constant
\begin{equation}\label{68-0}
     \gamma_0 =\frac{E_c}{2}\sin^2 \theta_0
     \tilde{\gamma}\left(\frac{d}{\ell}\right),
\end{equation}
where
\begin{equation}\label{68}
  \tilde{\gamma}(x)=
\begin{cases}x-\sqrt{\frac{\pi}{2}}\left(1-
e^{\frac{x^2}{2}}{\rm
erfc}\frac{x}{\sqrt{2}}\right)& \textrm{for}\ \lambda=0\\
x+\frac{x(3-x^2)}{16} -\sqrt{\frac{\pi}{2}}\left(\frac{11}{16}-
\frac{11-2 x^2+x^4}{16}e^{\frac{x^2}{2}}{\rm
erfc}\frac{x}{\sqrt{2}}\right)& \textrm{for}\ |\lambda|= 1
\end{cases}
\end{equation}
The quantity (\ref{68-0}) determines the pair interaction energy
per pair. The function $\gamma_0(d/\ell)$ is presented in Fig.
\ref{f2}. Evidently, the interaction energy for $\lambda={\pm 1}$
is greater than that for $\lambda=0$. Moreover, for small
$\tilde{d}$ the energy $\gamma_0 \propto d$ for $\lambda={\pm 1}$,
while for $\lambda=0$ the function $\gamma_0(d)$ is quadratic.

\begin{figure}
\begin{center}
\includegraphics[width=6cm]{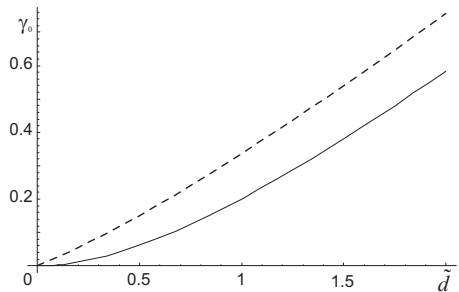}
\end{center}
\caption{Average interaction energy between pairs (in the units
$E_c \sin^2 \theta_0$) versus the distance between the layers.
Solid line -- $\lambda=0$, dashed line -- $|\lambda|=1$.}
\label{f2}
\end{figure}

In the long-wavelength limit the spectrum is linear: $E=\hbar s q
$ ($Q=0$), where the sound velocity equals
\begin{equation}\label{69}
s=\sqrt{\gamma_0 /M}= \frac{e^2}{\epsilon \hbar}\sin \theta_0
\tilde{s}\left(\frac{d}{\ell}\right),
\end{equation}
$${s}(x)=\frac{1}{2}\sqrt{\tilde{\gamma}(x)\tilde{\rho}(x)}.$$
A plot of the function ${s}(d/\ell)$ is presented in Fig.
\ref{f3}. According to Fig. \ref{f3}, the maximum sound velocity
(which is reached at $d/\ell\approx 1$ and zero imbalance of the
active component) does not depend on the magnetic field and for
$\epsilon\approx 3.9$ (SiO$_2$) it is $\approx 10^7$ cm/ s.

\begin{figure}
\begin{center}
\includegraphics[width=6cm]{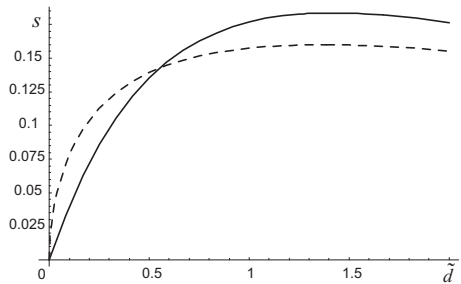}
\end{center}
\caption{Velocity of the acoustic mode (in the units
$e^2/(\varepsilon\hbar) \sin \theta_0$) versus the distance
between the layers. Solid line -- $\lambda=0$; dashed line --
$|\lambda|=1$.} \label{f3}
\end{figure}

The expression (\ref{40-3}) in the limits $q\to 0$, $Q\to 0$
assumes the form $v=(\hbar Q/M) \cos \theta_0$. Since $Q=\partial
\varphi/\partial X$, the quantity $v$ is the product of the
superfluid velocity ($v_s=\hbar \nabla\varphi/M$) and an
additional factor that depends on the imbalance. Specifically, for
zero imbalance $v=0$. As discussed in Ref. \cite{40}, this is a
consequence of the electron-hole symmetry. In the present work we
want to elucidate this feature on the basis of several other
arguments.

The superfluidity of electron-hole pairs in a bilayer quantum Hall
system can be viewed as an analog of the so-called counterflow
superfluidity \cite{41}. Counterflow superfluidity arises in a
system in which type I and II bosons on a lattice are present, the
total filling factor of the lattice sites is 1, and strong
same-site repulsion forbids two bosons from occupying the same
site. An elementary act of boson motion in such a system is a
type-I boson hopping onto a site with a type-II boson and the
type-II boson simultaneously hopping onto the site occupied by the
type-I boson. The fluxes of type-I and -II bosons in such a system
are equal in modulus and are oppositely directed. The ratio of the
velocity of the components I and II depends on the concentration
ratios of the components. For equal concentrations the moduli of
the velocities are likewise the same.

In the bilayer quantum Hall system considered here the number of
electrons of the top layer is equal to the number of holes in the
bottom layer, and likewise the number of holes in the top layer is
equal to the number of electrons in the bottom layer. The pairing
of electrons and holes from neighboring layers can be regarded as
the formation of two kinds of pairs differing by the direction of
the dipole moment (upwards or downwards). In such an
interpretation pair motion is an exchange process between
different kinds of pairs. In other words, the flux of pairs of one
kind is accompanied by a counterflux of pairs of the other kind.
In the absence of imbalance the concentrations of different kinds
of pairs is the same.

Counterflow superfluidity is a particular case of a two-component
superfluid \cite{42,43,44}. In this particular case additional
conditions are imposed--equality of the counterfluxes and
conservation of the locally total density of the components (for
this reason there is only one and two oscillation modes). The
spectrum of the collective modes in the two-component system
\cite{43,44} possesses the feature that it does not contain terms
which are linear in the gradient of the phase, if the velocities
of the components are equal in magnitude and opposite in
direction. At the same time this spectrum contains a quadratic
dependence on the gradient of the phase. The spectrum (\ref{39-1})
demonstrates similar properties, and vanishing of $v$ with zero
imbalance is a consequence of the counterflow character of the
superfluidity in a bilayer quantum Hall system.

For finite values of $q$ the spectrum deviates considerably from
the Bogolyubov spectrum. Specifically, for sufficiently large $d$
a roton-like minimum appears in the spectrum; the depth of this
minimum increases with increasing $d$ (Fig. \ref{f4}). For $d>d_c$
the spectrum becomes imaginary for finite $q$ and an instability
relative to the formation of a charge density wave arises in the
system. The dependence of the critical distance $d_c$ between the
layers on the imbalance of the filling factors of the active
component is presented in Fig. \ref{f5}, whence it is evident that
the component with pairing of the electrons and holes on the
Landau level $\lambda=\pm 1$ is unstable for smaller $d$ than the
same state arising in a situation where the active level is
$\lambda=0$.

\begin{figure}
\begin{center}
\includegraphics[width=6cm]{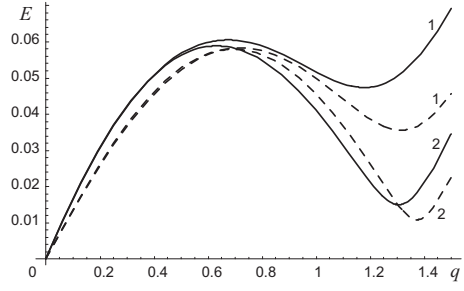}
\end{center}
\caption{Energy of the collective mode (in the units $E_c$) versus
the wave vector $q$ (in the units $\ell^{-1}$) for
$\tilde{\nu}_\beta=0$ and $d=0.9 d_c$ (1) and $d=0.99 d_c$ (2).
Solid lines -- $\lambda=0$, dashed lines -- $|\lambda|=1$.}
\label{f4}
\end{figure}

\begin{figure}
\begin{center}
\includegraphics[width=6cm]{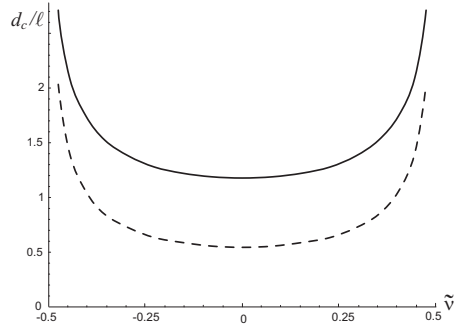}
\end{center}
\caption{Critical distance between layers versus the imbalance of
the filling factor of the active component. Solid line --
$\lambda=0$, dashed line  -- $|\lambda|=1$.} \label{f5}
\end{figure}

The critical values of the supercurrents can likewise be
determined by analyzing the spectrum (\ref{39-1}). As already
discussed in Ref. \cite{40}, the limitation on the currents in the
present case is not associated with the magnetic field which the
currents in the layers generate and is oriented parallel to the
layers. The maximum value of the current corresponds to the
maximum value of $Q$ for which the spectrum (\ref{39-1}) is
positive (Landau's superfluidity criterion) and real (stability
criterion). The dependence of the densities of the supercurrents
in the layers on $Q$ is given by the expression \cite{40}
\begin{equation} \label{cur}
    j_1 =-j_2 = \frac {e} {\hbar} \frac {1} {4\pi l^2} \sin^2 \theta_0 \frac {d
    \mathcal {F} _\mathrm{D} (Q)} {d Q}.
\end{equation}
The computational results for the critical current are presented
in Fig. \ref{f6}. The maximum critical current is proportional to
the quantizing magnetic field. For $B=1$ T and $\varepsilon=3.9$
the maximum critical current density equals approximately $\approx
1$ A/m.

\begin{figure}
\begin{center}
\includegraphics[width=12cm]{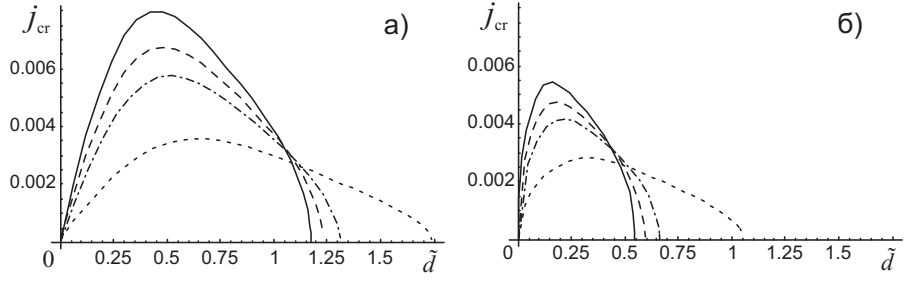}
\end{center}
\caption{Critical current density (in the units
$e^3/\hbar\varepsilon\ell^2$) versus the distance between the
layers for different magnitudes of the imbalance. Solid line --
$\tilde{\nu}=0$; dashed line -- $\tilde{\nu}=0.17$; dot-dash line
-- $\tilde{\nu}=0.25$; dashed line -- $\tilde{\nu}=0.4$ for
$\lambda=0$ (a) and $|\lambda|=1$ (b).} \label{f6}
\end{figure}

In closing this section we shall estimate the dependence of the
temperature of the transition into the superfluid state on $d$.
Since the system considered here is two-dimensional, the
transition into the superfluid state is a
Berezinskii-Kosterlitz-Thouless transition. The critical
transition temperature $T_c$ is determined by the equation
$T_c=\pi\rho_s(T_c)/2$, where $\rho_s(T)/2$ is the superfluid
stiffness at finite temperature. To find $\rho_s(T)$ we note that
the superfluid flux density $j_s$ is related with the gradient of
the phase $Q$ by the relation $j_s=\rho_s(T) Q/\hbar$ (for small
$Q$). On the other hand the flux density can be found from the
relation $j_s=(1/S \hbar)\partial F/\partial Q$, where
\begin{equation}\label{100}
    F=E_\textrm{mf}+E_\textrm{zp}+T\sum_{{\bf q}} \ln
    \left(1-\exp\left(-\frac{E({\bf q})}{T}\right)\right) \ -
\end{equation}
is the free energy. Here $E_\textrm{zp}$ is the energy of
zero-point vibrations. The collective and single-particle
excitations must be taken into account in the energy of the
zero-point vibrations. As shown in Ref. \cite{8}, for $d$ not too
close to the critical value the contribution of zero-point
vibrations to the renormalization of the superfluid stiffness is
negligible. Consequently, we shall neglect $E_\textrm{zp}$ when
calculating the superfluid stiffness. To take account of the
temperature correction to $\rho_s$ it is sufficient to include
only the collective modes in the entropy term in Eq. (\ref{100})
(the single-particle excitation spectrum has a gap of the order of
$E_c$). As a result we have
\begin{equation}\label{101}
    \rho_s(T)=\rho_{s0}+\frac{1}{S}\lim_{Q\to 0}\frac{1}{Q}\sum_{{\bf q}}\frac{\partial E({\bf
    q})}{\partial
    Q}N_B\left(E({\bf q})\right),
\end{equation}
where $ N_B(E)=(\exp({E}/{T})-1)^{-1}$ is the Bose distribution
function. For a single-component superfluid system with the
dispersion law $E({\bf q})=E_0(q)+\hbar q_x v_s$ (where $v_s=\hbar
Q/M$ is the superfluid velocity, and $E_0(q)$ is the spectrum as
$v_s=0$) the expression (\ref{101}) reduces to the standard form
$\rho_s(T)=\rho_{s0}-\rho_n$, where
\begin{equation}\label{102}
    \rho_n=-\frac{\hbar^2}{M S}\sum_{\bf q}\frac{\hbar^2
    q_x^2}{M}\left(\frac{\partial N_B(E)}{\partial
    E}\right)_{E=E_0(q)}\
\end{equation}
is the normal density. In our case the expression for the normal
density will be different, since the superfluidity is of a
counterflow character.

Here we shall focus on the calculation of the transition
temperature with zero imbalance. In this case the spectrum for
small $Q$ can be written in the form
$$E({\bf
q})=E_0(q)+\frac{1}{2}\alpha({\bf q}) Q^2+\ldots
$$
which gives
\begin{equation}\label{103}
    \rho_s(T)=\rho_{s0}+\frac{1}{S}\sum_{{\bf q}}\alpha({\bf
    q})N_B\left(E_0(q)\right).
\end{equation}

To calculate the coefficient $\alpha({\bf q})$ in the expansion we
shall require the spectrum of excitations for an arbitrary
direction of the wave vector. A method for calculating the desired
spectrum, generalizing the approach used in Ref. \cite{40}, is
described in the Appendix. According to the results obtained in
the Appendix, for $\theta_0=\pi/2$ the spectrum assumes the form
\begin{equation} \label{39-1-1}
E ({\bf q}) = \sqrt {\epsilon_{{\bf q},Q}  (\epsilon_{{\bf q},Q}
+2 \gamma_{{\bf q},Q})} ,
\end{equation}
where
\begin{equation} \label{40-1-1} \epsilon_{{\bf q},Q} =2\mathcal {F} _{D}
(Q) - \mathcal {F} _\textrm{D} (|{\bf q}+Q \hat{x}|) - \mathcal
{F} _\textrm{D} (|{\bf q}- Q \hat{x}|),
\end{equation}
\begin{equation} \label{40-2-1}
\gamma_{{\bf q},Q}  =  \left [\mathcal {H} ({\bf q},Q) - \mathcal
{F} _\textrm{S} (q) + \frac{\mathcal {F} _\textrm{D} (|{\bf q}+ Q
\hat{x}|)+ \mathcal {F} _\textrm{D} (|{\bf q}- Q
\hat{x}|)}{2}\right],
\end{equation}
the functions $\mathcal {F} _\textrm{S}(q)$ and $\mathcal {F}
_\textrm{D}(q)$ are determined by Eq. (\ref{63}) and (\ref{64}),
$\hat{x}$ is a unit vector in the direction $x$, and
\begin{equation}\label{40-2-2}
\mathcal {H} ({\bf q},Q)=\frac {e^2} {2 \varepsilon l^2} \; e ^
{-\frac {q^2 l^2} {2}}
    \frac {1 - e ^ {-d |q |}\cos(q_y Q \ell^2)} {|\, q |}
    f_\lambda(q\ell).
\end{equation}

Using Eqs. (\ref{39-1-1}-\ref{40-2-2}) to calculate the function
$\alpha({\bf q})$ we arrive are the following expression for
$\rho_s(T)$:
\begin{equation}\label{107}
    \rho_s(T)=\rho_{s0}+\int_0^{\infty} d q\frac{q
    N_B(E_0(q))
    }{2\pi E_0(q)}\left(2(\gamma_q+\epsilon_q)\mathcal {F} _\textrm{D}^{''}(0)-
    \left(\gamma_q+\frac{\epsilon_q}{2}\right) \left(\mathcal {F}
_\textrm{D}^{''}(q) + \frac{\mathcal {F}
_\textrm{D}^{'}(q)}{q}\right)+\epsilon_q \mathcal {D}(q)\right),
\end{equation}
where $$\mathcal {D}(q)=\frac {e^2} {4 \varepsilon } \; e ^
{-\frac {q^2 l^2} {2}}
     q \ell^2 e ^ {-d q }  f_\lambda(q\ell),
$$
$\epsilon_q=\epsilon_{{\bf q},0}$ and $\gamma_q=\gamma_{{\bf
q},0}$. The dependence of the critical temperature on the distance
between the layers, as calculated from Eq. (\ref{107}), is
presented in Fig. \ref{f7}.

\begin{figure}
\begin{center}
\includegraphics[width=6cm]{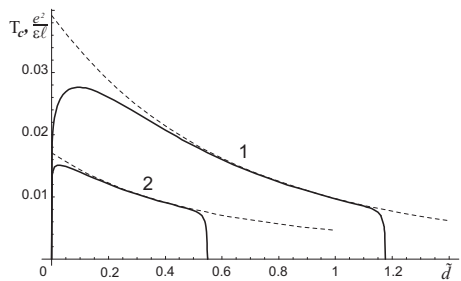}
\end{center}
\caption{Critical temperature versus the distance between the
layers. The solid curves 1 and 2 correspond to $\lambda=0$ and
$|\lambda|=1$, respectively. The dashed curves show the critical
temperature in the mean-field approximation (neglecting the
temperature renormalization of the superfluid stiffness).}
\label{f7}
\end{figure}

According to the results obtained, for intermediate values of $d$
the expression $T_c=\pi \rho_{s0}/2$ is a good approximation for
the transition temperature; it is important to take account of the
temperature correction for small $d/\ell$ and for $d$ close to the
critical values. The maximum critical temperature is attained for
intermediate values of $d$. The dependence $T_c(d/\ell)$ is
qualitatively similar to the dependence of the critical
temperature on $d/\ell$. In a state where electrons and holes from
the zeroth Landau level form pairs the critical temperature is
higher than in a state where electrons in the level $\lambda=+1$
form pairs with holes in the level $\lambda=-1$.

We note that the results of the present section, which were
obtained for the zeroth level, also describe the situation in
bilayer quantum Hall systems in semiconductor heterostructures
based on GaAs (and, correspondingly, partially repeat the results
of Refs. \cite{6,7,8,10,40}).

\section{Discussion}

The essential result following from the preceding section is that
the maximum parameters of electron-hole superconductivity in a
bilayer graphene system in a quantizing magnetic field can be
attained if interlayer phase coherence arises between electrons
belonging to the zeroth Landau level. In this connection there
arises the question of whether or not there are any advantages to
using graphene instead of GaAs heterostructures.

To answer this question we recall the general conditions under
which a superfluid state of bound electron-hole pairs can be
expected to appear in a bilayer quantum Hall system. Three
characteristic energies can be identified in the present problem:
the splitting $\omega_0$ between the Landau levels, the Coulomb
energy $E_c$, and the interlayer tunneling amplitude $t$. The
analysis performed in the present work is based on the assumption
that the following strong inequalities hold: $t\ll E_c$ and
$E_c\ll \omega_0$.

The first inequality make is possible to neglect tunneling when
analyzing collective modes. In addition, even weak tunneling
results in the formation of a system of vortices which are similar
to Josephson vortices. If a bilayer system is used to transfer
current from source to load (this is the configuration realized in
Refs. \cite{20,22,23}), the vortices will move, which will result
in energy dissipation \cite{45} whose magnitude dissipation is
proportional to the squared tunneling amplitude. We note that a
configuration where the vortices are stationary and dissipation is
absent can be realized \cite{46} but such configurations cannot be
used to transfer energy from source to load \cite{47}.
Consequently, systems with a negligibly small interlayer tunneling
amplitude are needed. In a double quantum well the tunneling is
quite large because of the relative narrowness and smallness of
the barrier.

The second inequality is the condition for using the so-called
lowest Landau level approximation (where transitions to inactive
levels are neglected). In this case the electrons and holes in the
active level are well-determined quasiparticles. In systems with a
quadratic dispersion law for charge carries the splitting between
the Landau levels $\omega_0=\hbar e B/m_* c$ ($m_*$ is the
effective mass of the carriers) is directly proportional to the
magnetic field and the Coulomb energy
$E_c=e^2/\varepsilon\ell=(e^2/\varepsilon)\sqrt{eB/\hbar c}$ is
proportional to the square root of the magnetic field.
Consequently, the inequality $E_c<\omega_0$ holds only in quite
strong fields ($B\gtrsim 10$ T for GaAs). The magnetic length is
small, i.e. the critical value of $d$ is small. Correspondingly,
large values of $d$ cannot be used to suppress interlayer
tunneling.

For a bilayer graphene system $\omega_0=\sqrt{2}\hbar v_F/\ell$,
i.e. the condition $E_c<\omega_c$ is equivalent to the condition
$\varepsilon>e^2/(\sqrt{2}\hbar v_F)\approx 1.5$. For an
appropriate choice of $\varepsilon$ the condition is satisfied for
arbitrary magnetic fields and large values of $d$ can be used (the
ratio $d/\ell$ can be made less than 1 by an appropriate choice of
the magnetic length). Since graphene is a monolayer and not a
sufficiently wide quantum well, the interlayer tunneling amplitude
will be much less than for the same values of $d$.

We note that the inequality $E_c\ll \omega_0$ is not so
fundamental for a graphene system as for a bilayer electronic
system in a GaAs heterostructure. In the second case, in the limit
of weak magnetic fields we switch to the case of two layers with
carriers of the same type, and in the first case (in a gate field)
we arrive at an electron-hole graphene system in a zero magnetic
field \cite{27,28,29,30}. For $\varepsilon=1$ and when a large
number of positive Landau levels are filled in a single layer (and
the same number of empty negative layers in the other layer) the
approach used in the present work, strictly speaking, is not
applicable. Nonetheless, the pairing effect apparently remains.
The question of the transition temperature in this case remains
open for the time being.

In closing, we shall discuss the question of which maximum
superconducting transition temperatures can be reached in a
bilayer quantum Hall system based on graphene. Let us assume that
we have created a system with a prescribed value of $d$ and we
start to vary the magnetic field (at the same time adjusting the
gate field so as to remain in the regime with zero imbalance of
the active component). To determine the maximum transition
temperature we shall represent the plot shown in Fig. \ref{f7} in
a somewhat different form; specifically, we shall choose the
quantity $e^2/\varepsilon d$ as the unit of temperature. The
results are displayed in Fig. \ref{f8}. As one can see from the
dependences presented, the maximum temperature is $T_c\approx 0.01
e^2/\varepsilon d$ and is reached for $d/\ell=0.5\div 1$ in the
case where carriers in the zeroth level participate in the
pairing. For $d=50 \AA$ and $\varepsilon =3.9$ the maximum
temperature $T_c\approx 8$ K. The required magnetic fields
$B\gtrsim 25$ T. A transition temperature of the order of 1 K can
be attained with $d\approx 400 \AA$ (for the same $\varepsilon$)
in sufficiently strong fields $B\gtrsim 0.4$ T.

\begin{figure}
\begin{center}
\includegraphics[width=6cm]{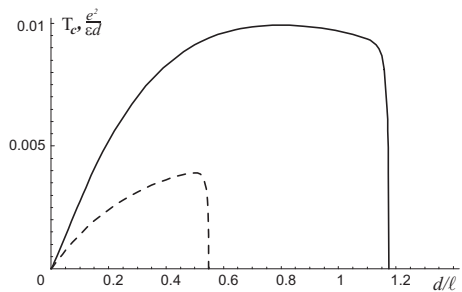}
\end{center}
\caption{Critical temperature with fixed distance between the
layers versus the reciprocal of the magnetic length. Solid line --
$\lambda=0$; dashed line -- $|\lambda|=1$.} \label{f8}
\end{figure}

\section*{APPENDIX. Spectrum of collective
oscillations with arbitrary direction of the wave vector}

\setcounter{equation}{0}
\def\theequation{A.\arabic{equation}}

The problem of the collective modes in a bilayer quantum system
with interlayer phase coherence is examined in Refs. \cite{8,9,10}
and \cite{40}. In Ref. \cite{40} the approach of Ref. \cite{10}
(based on the quantization of the energy of the phase fluctuations
and local imbalance) is extended to the case of an arbitrary
average imbalance. It is shown in \cite{40}  that the spectrum
possesses electron-hole symmetry and becomes a Bogolyubov spectrum
in the low-density limit. The case where the direction of the wave
vector coincides with the direction of the phase gradient has been
examined in Refs. \cite{10} and \cite{40}. In Refs. \cite{8} and
\cite{9} the functional integral approach is used to find the
spectrum. In Ref. \cite{8} the case of zero imbalance was examined
and the spectrum was found for an arbitrary direction of the wave
vector. The general case is studied in Ref. \cite{9}. In the
present Appendix we extend the approach used in Ref. \cite{40} to
the case of an arbitrary direction of the wave vector.

Using the analogy with a quantum ferromagnet [6] we shall examine
the Fourier components of the pseudospin density operators
\begin{equation}\label{a1}
    \hat{{\bf m}}({\bf q})=\sum_{k_1, k_2}\langle k_1|e^{- i {\bf q
    r}}|k_2\rangle a^+_{i,k_1} {\bm \sigma}_{ij}a_{j,k_2}
\end{equation}
where ${\bm \sigma}_{ij}$ are Pauli matrices. The state with
nonzero phase gradient can be interpreted as a helical state with
$m_z= \cos \theta_0$, $m_x=\sin \theta_0 \cos({\bf Q r})$,
$m_y=\sin \theta_0 \sin({\bf Q r})$. In such a state the Fourier
series expansion of the averages $\langle \Psi |\hat{
m}_{x(y)}({\bf q})|\Psi\rangle$ contains terms with ${\bf q}=\pm
{\bf Q}$.

The wave function
\begin{equation}\label{a2}
 |\Psi\rangle =
 \prod_{k}\left(\cos\frac{\theta_0}{2}
 a^+_{1,k}+ e^{i Q k}\sin \frac{\theta_0}{2}
 a^+_{2,k}\right)|0\rangle
\end{equation}
describes the helical state with ${\bf Q}=Q \hat{x}$, where
$\hat{x}$ is a unit vector along the $x$ axis. To describe the
helical state with arbitrary ${\bf Q}=(Q_x, Q_y)$ the wave
function must be chosen in the form
\begin{equation}\label{a3}
 |\Psi\rangle =
 \prod_{k}\left(\cos\frac{\theta_0}{2}
 a^+_{1,k}+ e^{i Q_x k}\sin \frac{\theta_0}{2}
 a^+_{2,k-Q_y}\right)|0\rangle.
\end{equation}
It is easily verified that the energy is independent of the
direction of ${\bf Q}$: in the state (\ref{a3}), just as in the
state (\ref{a2}), the expression for the energy has the form
(\ref{65}). Indeed, the transition from (\ref{a2}) to (\ref{a3})
is simply a rotation of the coordinate system, and the energy
should not change under such a transformation. We shall examine
the fluctuations of $\theta$ and $\varphi$ along the $x$ axis in a
new coordinate system. To take account of such fluctuations we
write the wave function in the form
\begin{equation}\label{a4}
 |\Psi\rangle =
 \prod_{k}\left(\cos\frac{\theta_k}{2}
 a^+_{1,k}+ e^{i Q_x k+i \tilde{\varphi}_k}\sin \frac{\theta_k}{2}
 a^+_{2,k-Q_y}\right)|0\rangle.
\end{equation}
A calculation of the energy in the state (\ref{a4}) gives
\begin{eqnarray} \label{a5}
 E =E_0 -{e\tilde{V}} \sum_{X}  \cos \theta_X+\frac {1} {2 L_\mathrm{y}}
 \sum _ {X, X '} \Big \{\left [H_{Q_y}
(X-X ') - F_\textrm{S} (X-X ') \right] \cos \theta_X \cos \theta _
{X '} \cr
 - F_{\textrm{D},Q_y} (X-X ') \sin\theta_X \sin\theta _ {X '}
\cos (Q_x(X-X')+\tilde{\varphi}_X - \tilde{\varphi} _ {X '}) \Big
\},
\end{eqnarray}
where
\begin{eqnarray} \label{a6}
H_{Q_y}(X) = \frac {e^2} {2 \varepsilon} \int _ {-\infty} ^
{\infty} d q \frac {1-e ^ {-|q | d} \cos(Q_y q\ell^2)} {|\, q |}
\; e ^ {{i} q X - \frac {q^2 l^2} {2}}f_\lambda(q\ell), \cr
F_\mathrm{S} (X) = \frac {e^2} {2 \varepsilon} \; e ^ {-\frac
{X^2} {2 l^2}} \int _ {-\infty} ^ {\infty} \frac {d q} {\sqrt {q^2
+ X^2/l^4}} \; e ^ {- \frac {q^2 l^2}
{2}}f_\lambda\left(\sqrt{q^2\ell^2+X^2/\ell^2}\right), \cr
F_{\mathrm{D}, Q_y} (X) = \frac {e^2} {2 \varepsilon} \; e ^
{-\frac {X^2} {2 l^2}} \int _ {-\infty} ^ {\infty} \frac {d q}
{\sqrt {q^2 + X^2/l^4}} \; e ^ {-d\sqrt{q^2 + X^2/l^4}  - \frac
{q^2 l^2} {2}} e^{-i q_x
Q_y}f_\lambda\left(\sqrt{q^2\ell^2+X^2/\ell^2}\right).
\end{eqnarray}

Performing the expansion (\ref{a6}) with respect to small
fluctuations $\tilde{\varphi}_X$ and $\tilde {m} _z (X)= \cos
\theta_X-\cos \theta_0$ and switching to the Fourier components
\begin{equation} \label{a7}
\tilde{m} _ {z} (q) = \frac {2 \pi l^2} {S} \sum_X \tilde{m} _ {z}
(X) e ^ {-{i} q X}, \quad \tilde{\varphi} (q) = \frac {2 \pi l^2}
{S} \sum_X \tilde{\varphi} (X) e ^ {-{i} q X},
\end{equation}
we obtain the following expression for the energy of the
fluctuations:
\begin{equation} \label{a8}
 E_{{\rm fl}}= \frac {S} {4 \pi l^2} \sum_q \Big [\tilde{m} _ {z}
(-q) \mathcal {K} _ {zz} (q) \tilde{m} _ {z} (q)   +
\tilde{\varphi} (-q) \mathcal {K} _ {\varphi \varphi} (q)
\tilde{\varphi} (q) - \left(i\tilde{m}_z (-q) \mathcal {K} _ {z
\varphi} (q) \tilde{\varphi} (q)+c.c.\right)\Big],
\end{equation}
where
\begin{equation} \label{a9}
 \mathcal {K} _ {zz} (q) = \mathcal {H}_{Q_y} (q) - \mathcal {F}
_\mathrm{S} (q) + \mathcal {F} _\mathrm{D} (Q)  + \left (\mathcal
{F} _\mathrm{D} (Q) - \frac {\mathcal {F} _\mathrm{D} (|q
\hat{x}+{\bf Q}|) + \mathcal {F} _\mathrm{D} (|q\hat{x} - {\bf
Q})|} {2} \right) \cot^2 \theta_0,
\end{equation}
\begin{equation} \label{a10}
\mathcal {K} _ {z \varphi}(q) = \cos \theta_0  \frac {\mathcal {F}
_\mathrm{D} (|q \hat{x}+{\bf Q}|) - \mathcal {F} _\mathrm{D} (|q
\hat{x}-{\bf Q}|)} {2},
\end{equation}
\begin{equation} \label{a11}
\mathcal {K} _ {\varphi \varphi} (q) =  {\sin^2 \theta_0}  \left
[\mathcal {F} _\mathrm{D} (Q) - \frac {\mathcal {F} _\mathrm{D}
(|q \hat{x}+{\bf Q}|) + \mathcal {F} _\mathrm{D} (|q \hat{x}-{\bf
Q}|)} {2} \right].
\end{equation}
\begin{equation} \label{a12}
    \mathcal {H}_{Q_y} (q) = \frac {e^2} {2 \varepsilon l^2} \; e ^ {-\frac {q^2 l^2} {2}}
    \frac {1 - e ^ {-d |q |}\cos(Q_y q\ell^2)} {|\, q |} f_\lambda(q\ell),
\end{equation}
and the functions $\mathcal {F} _\mathrm{S,D}(q)$ are determined
by Eqs. (\ref{63}) and (\ref{64}). To quantize the energy
(\ref{a8}) we take account of the fact that $\tilde {m} _z $ and $
\tilde{\varphi}$ are canonically conjugate and the commutator of
the operators of these quantities equals
\begin{equation} \label{a13}
[\hat {m} _ {z} (q), \hat {\varphi} (q ')] = - 2 {i} \frac {2 \pi
l^2} {S} \delta _ {q,-q '}.
\end{equation}
Expressing the operators $\hat{m} _ {z} (q) $ and $ \hat{\varphi}
(q) $ in terms of Bose creation and annihilation operators
\begin{equation} \label{a14}
\hat {m} _ {z} (q) = A_q \sqrt{\frac{2 \pi \ell^2}{S}}( {b} _ {q}
+ {b} ^ + _ {-q}), \ \hat {\varphi} (q) = {i} \frac{1}{A_q}
\sqrt{\frac{2 \pi \ell^2}{S}}( {b} _ {q} - {b} ^ + _ {-q}),
\end{equation}
replacing in Eq. (\ref{a8}) the quantities $\tilde {m} _z $ and $
\tilde{\varphi}$ by the operators (\ref{a14}), and requiring the
terms containing two creation or two annihilation operators to
vanish (which makes it possible to determine $A_q$), we arrive at
the equation
\begin{equation} \label{a15}
H_{\mathrm{fl}} =  \sum _ {q} E (q) \left (b ^ + (q) b (q) + \frac
{1} {2} \right),
\end{equation}
where
\begin{equation} \label{a16}
E (q) = 2 \left (\sqrt { \mathcal {K} _ {\varphi \varphi}
(q)\mathcal {K} _ {zz} (q)} + {\mathcal {K}} _ {z \varphi} (q)
\right).
\end{equation}
The spectrum (\ref{a16}) of the collective modes depends on $q$,
$Q$, and the angle between the $x$ axis (which is chosen in the
direction of the wave vector ${\bf q}$) and the vector ${\bf Q}$.
Rotating the coordinate system to that the $x$ axis is oriented
along ${\bf Q}$, we obtain the spectrum of the collective modes
for an arbitrary direction of the wave vector:
\begin{equation} \label{a17}
E ({\bf q}) = \sqrt {\epsilon_{{\bf q},Q}  (\epsilon_{{\bf q},Q}
+2 \gamma_{{\bf q},Q}\sin^2 \theta_0)} + \cos \theta_0
\left(\mathcal {F} _\mathrm{D} (|{\bf q}+Q\hat{x}|) - \mathcal {F}
_\mathrm{D} (|{\bf q}- Q\hat{x}|)\right),
\end{equation}
where the quantities $\epsilon_{{\bf q},Q}$ and $\gamma_{{\bf
q},Q}$ are determined by the expressions (\ref{40-1-1}) and
(\ref{40-2-1}), respectively.

\end{document}